\pdfoutput=1
\documentclass[twoside,twocolumn,9pt]{article}
\usepackage{extsizes}
\usepackage[super,sort&compress,comma]{natbib} 
\usepackage[version=3]{mhchem}
\usepackage[left=1.5cm, right=1.5cm, top=1.785cm, bottom=2.0cm]{geometry}
\usepackage{balance}
\usepackage{mathptmx}
\usepackage{sectsty}
\usepackage{graphicx} 
\usepackage{lastpage}
\usepackage[format=plain,justification=justified,singlelinecheck=false,font={stretch=1.125,small,sf},labelfont=bf,labelsep=space]{caption}
\usepackage{float}
\usepackage{fancyhdr}
\usepackage{fnpos}
\usepackage[english]{babel}
\addto{\captionsenglish}{%
  
}
\usepackage{array}
\usepackage{droidsans}
\usepackage{charter}
\usepackage[T1]{fontenc}
\usepackage[usenames,dvipsnames]{xcolor}
\usepackage{setspace}
\usepackage[compact]{titlesec}
\usepackage{hyperref}

\usepackage{epstopdf}

\definecolor{cream}{RGB}{222,217,201}

\begin{document}

\pagestyle{fancy}
\thispagestyle{plain}
\fancypagestyle{plain}{
\renewcommand{\headrulewidth}{0pt}
}

\makeFNbottom
\makeatletter
\renewcommand\LARGE{\@setfontsize\LARGE{15pt}{17}}
\renewcommand\Large{\@setfontsize\Large{12pt}{14}}
\renewcommand\large{\@setfontsize\large{10pt}{12}}
\renewcommand\footnotesize{\@setfontsize\footnotesize{7pt}{10}}
\makeatother

\renewcommand{\thefootnote}{\fnsymbol{footnote}}
\renewcommand\footnoterule{\vspace*{1pt}%
\color{cream}\hrule width 3.5in height 0.4pt \color{black}\vspace*{5pt}} 
\setcounter{secnumdepth}{5}

\makeatletter 
\renewcommand\@biblabel[1]{#1}            
\renewcommand\@makefntext[1]%
{\noindent\makebox[0pt][r]{\@thefnmark\,}#1}
\makeatother 
\renewcommand{\figurename}{\small{Fig.}~}
\sectionfont{\sffamily\Large}
\subsectionfont{\normalsize}
\subsubsectionfont{\bf}
\setstretch{1.125} 
\setlength{\skip\footins}{0.8cm}
\setlength{\footnotesep}{0.25cm}
\setlength{\jot}{10pt}
\titlespacing*{\section}{0pt}{4pt}{4pt}
\titlespacing*{\subsection}{0pt}{15pt}{1pt}

\fancyfoot{}
\fancyfoot[LO,RE]{\vspace{-7.1pt}\includegraphics[height=9pt]{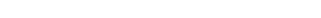}}
\fancyfoot[CO]{\vspace{-7.1pt}\hspace{13.2cm}}
\fancyfoot[CE]{\vspace{-7.2pt}}
\fancyfoot[RO]{\footnotesize{\sffamily{ ~\textbar  \hspace{2pt}\thepage}}}
\fancyfoot[LE]{\footnotesize{\sffamily{\thepage~\textbar }}}
\fancyhead{}
\renewcommand{\headrulewidth}{0pt} 
\renewcommand{\footrulewidth}{0pt}
\setlength{\arrayrulewidth}{1pt}
\setlength{\columnsep}{6.5mm}
\setlength\bibsep{1pt}

\makeatletter 
\newlength{\figrulesep} 
\setlength{\figrulesep}{0.5\textfloatsep} 

\newcommand{\topfigrule}{\vspace*{-1pt}%
\noindent{\color{cream}\rule[-\figrulesep]{\columnwidth}{1.5pt}} }

\newcommand{\botfigrule}{\vspace*{-2pt}%
\noindent{\color{cream}\rule[\figrulesep]{\columnwidth}{1.5pt}} }

\newcommand{\dblfigrule}{\vspace*{-1pt}%
\noindent{\color{cream}\rule[-\figrulesep]{\textwidth}{1.5pt}} }

\makeatother

\twocolumn[
  \begin{@twocolumnfalse}
{
}
\sffamily

\begin{center}
\LARGE{\textbf{Structure, dynamics and phase transitions in electric field assembled colloidal crystals and glasses}} \\

\vspace{0.5cm} 

\large{Indira Barros,\textit{$^{a}$} Sayanth Ramachandran,\textit{$^{a, b}$} and Indrani Chakraborty$^{\ast}$\textit{$^{a}$}} \\
\end{center}

\normalsize{Field-induced assembly of colloidal particles into structures of desired configurations is extremely relevant from the viewpoint of producing field-assembled micro-swimmers and reconfigurable smart materials. However, the behaviour of colloidal particles under the influence of alternating current (AC) electric fields remains a topic of ongoing investigation due to the complex effects of various control parameters. Here we examine the role of several factors including particle size, zeta potential, voltage and frequency of the applied field in the formation of different structural configurations ranging from crystals to glasses, and observe interesting and unexpected behaviours in the structure formation. Additionally, we investigate the dynamics of structure formation; the nature of diffusion and active motion in these out-of-equilibrium systems, and show how that leads to the formation of close-packed or open structures. Lastly, we investigate the frequency-driven disorder-order-disorder phase transition in colloidal crystals, which is a starting point for building reconfigurable systems. Our findings contribute to a deeper understanding of the significant roles of various factors in electric field-induced assembly of colloidal particles, as well as pave the way for potential applications in micro-robotics and next generation materials.} \\


 \end{@twocolumnfalse} \vspace{0.6cm}

  ]

\renewcommand*\rmdefault{bch}\normalfont\upshape
\rmfamily
\section*{}
\vspace{-1cm}


\footnotetext{\textit{$^{a}$~Department of Physics, Birla Institute of Technology and Science, Pilani–K K Birla Goa Campus, Zuarinagar, Goa 403726, India, E-mail: indranic@goa.bits-pilani.ac.in}}
\footnotetext{\textit{$^{b}$~Max-Planck Institute for Polymer Research Mainz, Ackermannweg 10, 55128 Mainz, Germany. }}





\section{Introduction}
Programmable self-assembly of colloidal particles into desired configurations has recently become an area of intense interest due to its tremendous application potential in new-age manufacturing of smart, reconfigurable materials, construction of microbots for targeted drug delivery and fabrication of the next-generation biomedical devices. \cite{nelson2010microrobotsmedicine,ambarishrootcanal,biomedical,drugdelivery, reviewsmartmaterials}. Colloids are an excellent choice as building blocks in the micrometer and nanometer scales for a controlled assembly of several complex architectures, ranging from colloidal oligomers and reconfigurable `molecules' to colloidal crystals with different symmetries and structural configurations.  \cite{velev2009materialengineering, li2011colloidalassembly, chakraborty2017colloidaljoints, chakraborty2022colloidalmolecules,pine2015crystallization, pine2020colloidaldiamond, pine2020reconfigurable, sacanna2010lockkey, khalil2012ising} Several different interactions, ranging from capillary, \cite{wang2020capillary, liu2018capillary} depletion, \cite{sacanna2010lockkey} DNA mediated, \cite{chakraborty2022colloidalmolecules, chakraborty2017colloidaljoints, pine2015crystallization, wang2012naturebonding}to field induced \cite{liljestrom2019assemblyreview, harraq2022fieldlangmuir} interactions have been used to produce a wide variety of architectures using colloidal building blocks. Field-induced assembly, where colloids are assembled into structures of desired configurations using electric field, \cite{ningwucolloidalmolecules, ningwunonclosepacked, lee2019helical} magnetic field\cite{yang2020pnasreconfigurable, han2020microscallops} or light \cite{palacci2013lightlivingcrystals} induced activation, has also been studied extensively in the last few years \cite{liljestrom2019assemblyreview}. The major advantages of field-induced assembly are: a) the morphology of the produced structures can be made fully controllable and configurations can be switched from one to another by varying the field parameters, b) magnetic fields can penetrate live tissues, leading to assembly, control and propulsion of such a structure inside biological systems,\cite{rotundo2022magneticfieldbody} and c) electric field based activation is easy to set up, and has a low cost.

 Electric field-induced assembly offers remarkable versatility in assembling colloidal particles, thanks to the multitude of controllable parameters within the system. However, this versatility also entails a wide spectrum of phenomena and structural configurations achievable by small changes in the control parameters. Trau \textit{et al.} observed the formation of two-dimensional colloidal crystals on electrode surfaces for both micrometer and nanometer-sized particles upon the application of a DC electric field. \cite{trau1996field} They were also able to produce multilayers by controlling the DC field and by field-induced annealing with a low-frequency AC field. Using an out-of-plane AC electric field, Ma \textit{et al.} observed the formation of colloidal oligomers of different valences in the low-frequency regime (0.5-1.5 kHz) and the formation of honeycomb-like open structures at higher frequencies. \cite{ningwucolloidalmolecules} By superimposing a DC electric field with an AC electric field, Maestas \textit{et al.} controlled the number ratio of particles in different planes, thereby producing square and triangular bilayers, rectangular bands, zig-zagged stripes, sigma lattice and honeycomb-Kagome structures.\cite{maestas2021electric} Ordered, non-close packed colloidal arrays were produced by Jingjing \textit{et al.} using a combination of a low-frequency AC electric field and a sequence of DC pulses in a solution of appropriate particle concentrations. \cite{ningwunonclosepacked} Yakovlev \textit{et al.} made  two-dimensional crystalline structures with a rotating electric field \cite{yakovlev2017tunable}. Biaxial fields were used by Leunissen \textit{et al.} \cite{leunissen2009biaxial} to produce chains at moderate fields and one-particle thick sheet-like structures at higher fields. Moving one step further, Vissers \textit{et al.} showed a distinct separation of binary particle species with opposing charges into bands perpendicular to the applied AC electric field at low frequencies ($\sim$ 1.5 Hz). \cite{vissers2011band} Heatley \textit{et al.} produced colloidal molecules of tunable size and bond length using a binary mixture of colloidal particles of different types and sizes under a perpendicularly applied AC electric field.\cite{heatley2017colloidalmolecules} `Active colloidal molecules' utilizing binary colloids, patchy particles and even live cells were produced by Wang \textit{et al.} \cite{wang2020activecolloidalmolecules} Other than isotropic particles, anisotropic particles like colloidal dimers were observed to align perpendicular to the substrate to form close-packed crystals under an AC electric field.\cite{ma2012two} Using colloidal dimers actuated by a perpendicularly applied AC electric field, Yang \textit{et al.} obtained stable planar clusters with handedness, \cite{yang2018colloidalmotors} while Katzmeier \textit{et al.} used asymmetric dimer particles to demonstrate propulsion under AC electric fields for potential applications as micro-swimmers. \cite{katzmeier2023natcommicrorobots}
 \vspace*{-0.2mm}
 \\ \hspace*{0.1mm} In spite of the sheer diversity of phenomena in electric field-induced colloidal assembly, the difficulty of controlling the sensitivity to minor parameter variations and understanding the nuanced interplay of several influencing factors that determine the outcome of the final structure makes field-induced assembly challenging. This emphasizes the necessity for a detailed exploration of the various influencing parameters, and understanding the complex dynamics of structure formation. In our work, we systematically study the dependence of the electric field (AC) induced structure formation of isotropic colloids on various parameters such as the particle size, zeta potential, frequency and voltage of the applied electric field. We show that slight changes in parameters such as the zeta potential can lead to completely unexpected structure formations, in contrast to several previous results. Furthermore, we make a detailed study of the diffusion dynamics of the particles as they assemble into a structure. We show that the dynamics of motion of the particles is directly interlinked with the geometry and order of the structures produced. Thirdly, we probe the frequency-induced phase transition of the colloidal particles from an open glassy structure to an ordered crystalline structure and finally to a disordered, bulk structure. This is an extremely relevant attribute for building reconfigurable colloidal crystals, which have garnered a lot of interest over the last few years because of their potential use in switchable photonic crystals and smart materials for sensing. \cite{lumsdon2004photonic, HU2024103089, NAMIGATA2024100806} Additionally, field-assembled and field-driven micro-swimmers can be used for targeted drug delivery.\cite{katzmeier2023natcommicrorobots, pharmaceutics12070665} Therefore this study not only bridges the gap in our understanding of colloidal assembly and dynamics under electric fields but also helps in engineering complex structures with tailored functionalities.

\begin{figure}[!h]
\centering
  \includegraphics[width=8.3cm]{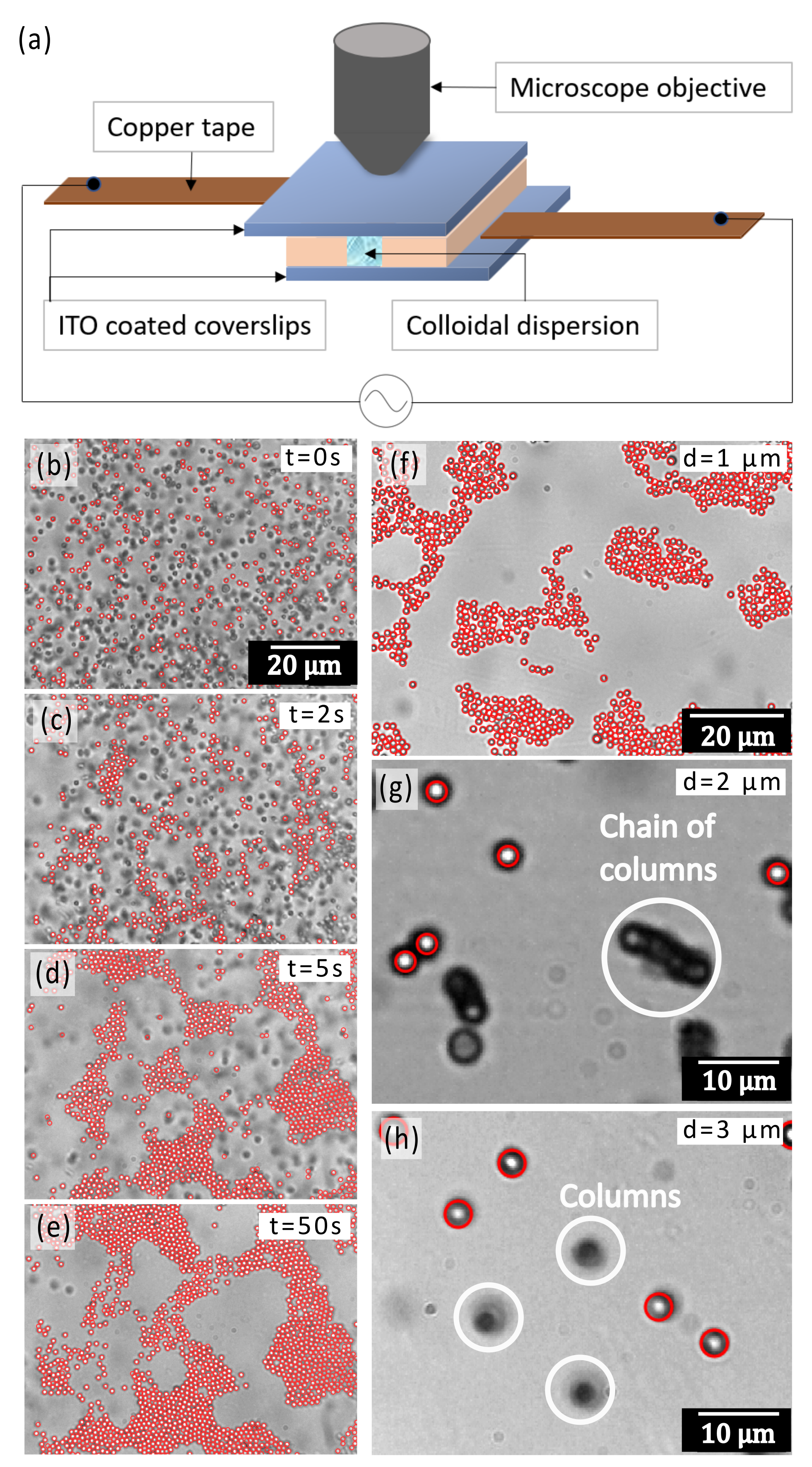}
  \caption{(a) Schematic of the sample chamber with the colloidal dispersion under an AC electric field applied perpendicular to the glass coverslips. (b-e) Time evolution images of 1 $\mu $m diameter particles (d= 1 $\mu $m) dispersed in a solution containing NaCl ($10^{-5}$ M) forming close-packed crystals when subjected to an electric field of frequency 5 kHz and peak-to-peak voltage of 20 V. (f) 1 $\mu$m particles (without salt) showing planar glassy structure formation, (g) 2 $\mu m$ particles forming chains of vertical columns and (h) 3 $\mu$m particles forming vertical columns parallel to the applied field.}
  \label{fig1}
\end{figure}

\section{Experimental Details}
\subsection{Sample preparation and Imaging}
Polystyrene colloidal particles of 1 $\mu$m size were purchased from Bangs Laboratories and Sigma-Aldrich. The particles had different functionalizations as given in Table \ref{tbl:zeta} as a result of which they had different zeta potential values. A suspension of the desired particles dispersed in de-ionized water was prepared with a solid concentration of 0.01\%. A small concentration (4.33 mM) of a surfactant SDS was added to ensure a smooth transfer to the capillary chamber. In the case where salt was added to the suspension, a solution of 0.01 mM NaCl and 4.33 mM SDS was prepared in de-ionized water and the dispersion of the particles was made in this solution.

\begin{table}[h]
\small
  \caption{\ List of samples and details}
  \label{tbl:zeta}
  \begin{tabular*}{0.48\textwidth}{@{\extracolsep{\fill}}lll}
    \hline
    Sample name & Details & Zeta Potential (mV) \\
    \hline 
    \\  
    P1X & 1$\mu$m polystyrene & $-$82.9 $\pm$ 1.0  \vspace{2mm} \\
    P1A &  1$\mu$m amine  & $-$94.6 $\pm$ 1.7 \\ & functionalized polystyrene  \vspace{2mm} \\
    Pr1X & 1$\mu$m polystyrene & $-$42.5 $\pm$ 1.7  \\ & spheres embedded \\ & with magnetic nanoparticles  \vspace{2mm}\\
    P1X + NaCl & P1X with $10^{-5}$ M NaCl & $-$102.8 $\pm$ 3.1\\
      \\
    \hline
  \end{tabular*}
\end{table}

To prevent non-specific interactions between the particles and the substrate, Indium Tin Oxide (ITO)-coated glass coverslips (thickness \#1) were cleaned by sequential ultrasonic treatment in isopropyl alcohol for 15 minutes, acetone for 15 minutes, and again in isopropyl alcohol for 15 minutes. The coverslips were then rinsed with de-ionized water and dried in an oven at 80°C.

A capillary chamber was fabricated by sandwiching two cleaned ITO coated coverslips which serve as the electrodes. The electrodes were placed in a slightly displaced manner (see Fig. \ref{fig1}a) to enable smooth contact fabrication. Spacing of 150 $\mu$m between the electrodes was achieved using a plastic spacer which also prevents shorting of the electrodes. Electrical contacts were made with copper tape and silver paste. The colloidal dispersion was then carefully pipetted into the capillary and all ends were sealed using UV-curable adhesive, ensuring the containment of the particle dispersion within the capillary chamber. AC electric field was applied using a function generator (METRAVI DDS-1010) in the sinusoidal mode. The experimental setup was observed under an optical microscope, utilizing a 100x objective lens in bright-field mode. Videos were recorded using a high-speed camera. The zeta potential measurements were made using a Particulate Systems - NanoPlus zeta sizer. The final zeta potential values were obtained after averaging over three separate measurements.

\subsection{Analysis}
Trackpy, \cite{trackpycode} a Python-based implementation of the Crocker-Grier algorithm \cite{crocker1996methods} was used to track the particle positions from the sequence of recorded images. In many of the figures of this paper the tracked particles are annotated with red circles for clarity. The trajectories of the particles were determined over thousands of frames and from the trajectories, mean square displacement (\textit{MSD}) vs lag time plots and displacement distributions were generated. Subsequently, the diffusion exponent values ($n$) were calculated from the \textit{MSD} analysis. To analyze the formed structure, Voronoi diagrams were constructed in Python from which the number of nearest neighbours, nearest-neighbour distances and areas of the Voronoi cells were calculated. 

\begin{figure}[!h]
\centering
  \includegraphics[width=8.3cm]{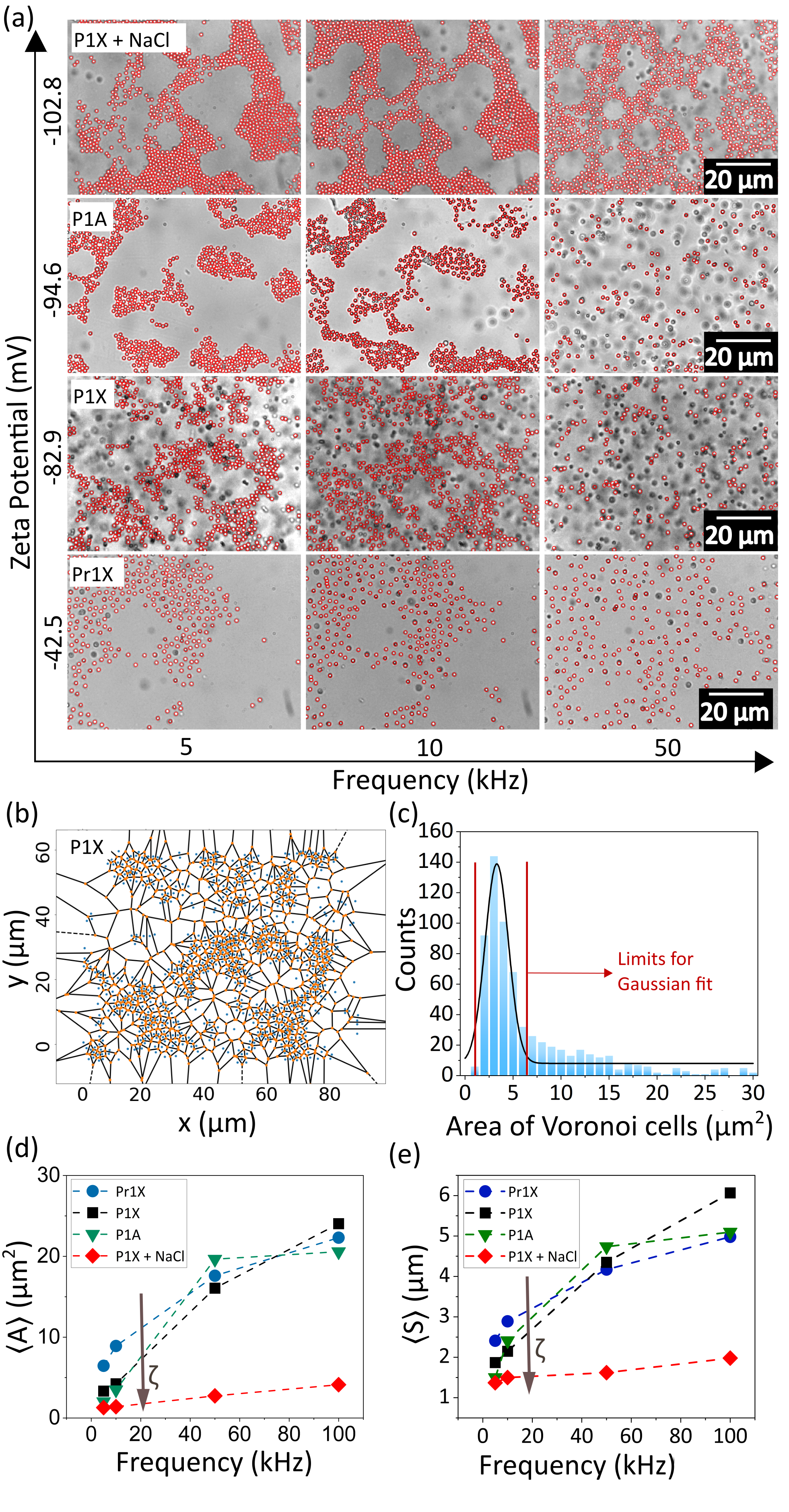}
  \caption{Comparison of the morphology of the colloidal clusters formed under different characteristic parameters. (a) A table of images showing the morphology of the structures formed with varied zeta potentials in response to a range of applied AC electric field frequencies. (b) A representative Voronoi diagram (sample P1X) at 5 kHz. (c) Gaussian fitting for the histogram of the area of the Voronoi cells as given in (b). (d) A plot of the mean Voronoi cell areas $(\langle A \rangle)$ vs frequency for samples with different $\zeta$ values all showing an overall increasing trend indicating the formation of more open structures with increasing frequency. $\langle A \rangle$ also increases with lower (negative) $\zeta $ values in the low-frequency regime ($\sim$ 5 kHz). (e) A plot of the average nearest neighbour distances $(\langle S \rangle)$ as a function of frequency also shows the same trend as in plot (d).}
  \label{fig2}
\end{figure}

\section{Results}
\subsection{Factors influencing structural configurations under AC electric field}
We studied the formation of colloidal crystals and glasses under AC electric fields, as a function of different structural and field parameters. The sample chamber was composed of two ITO-coated cover glasses with a spacer of 150 $\mu$m thickness (Fig. \ref{fig1}a). An AC electric field of desired peak-to-peak voltage (Vpp) and frequency was given in a direction perpendicular to the plane of the coverslips. A representative image of the formation of colloidal crystals composed of 1 $\mu$m polystyrene particles at a frequency of 5 kHz, Vpp= 20 V, and a salt concentration of $10^{-5}$ M as a function of time is shown in Fig.\ref{fig1}b-e. The crystals majorly show a hexagonal close-packed configuration with a packing fraction of 74\% in the inner regions. Over the span of 24 hours, the crystals formed could span large areas of the order of 100 $\times$ 100 $\mu$m\(^{2}\). Without the presence of any salt, we observed the formation of planar glassy structures with 1 $\mu$m diameter particles rather than ordered crystals (Fig. \ref{fig1}f). However, size played a major role in determining the final structure. In contrast to the planar glassy structures formed by the 1 $\mu$m polystyrene particles, the 3 $\mu$m diameter particles strongly preferred the formation of vertical columns along the direction of the applied electric field (Fig. \ref{fig1}h). The intermediate case was that of the 2 $\mu$m diameter particles, which showed simultaneous formation of columns and chains of these columns (Fig. \ref{fig1}g) indicating that the assembly occurs both in horizontal and vertical planes in this case.

The configuration of the assembled colloidal crystals and glasses was found to be strongly dependent on three parameters: a) applied voltage b) frequency of the AC field and c) zeta potential of the particles. The Zeta potential ($\zeta$) of the particles is obviously expected to play a major role in electric field-induced assembly of the particles. To examine this, we used a set of particles with different functionalizations and therefore different zeta potential values as given in Table \ref{tbl:zeta}. We tabulated the structures formed as a function of $\zeta$ and applied field frequency (Fig. \ref{fig2}a). In the low-frequency regime ($\sim$ 5 kHz), for the lowest value of the zeta potential in our experiments, that is $\zeta$ = $-$42.5 mV, we observed the formation of planar glassy structures with the highest nearest neighbour distances. In this case at higher frequencies, the particles covered the entire viewing plane with relatively fewer localized clusters. As the $\zeta$ increased to $-$82.9 mV, the nearest-neighbour distances between the particles started to decrease and at $\zeta$ = $-$94.6 mV, the formation of more packed configurations of localized clusters was observed. The highest value of $\zeta$ in our experiments was obtained with the addition of $10^{-5}$ M NaCl to sample P1X, which counter-intuitively increased the $\zeta$ value. This is due to the adsorption of ions on the surface of the colloids and the consequent increase in the surface charge as demonstrated by Manilo \textit{et al.}\cite{manilo2019salteffects} At the highest value of $\zeta$, the formation of close-packed nearly-crystalline structures was observed. This scenario changes with increasing values of the applied frequency. Firstly, with an increase in frequency, an overall increase of separation between the constituent particles of a cluster was observed, while lower frequencies (5-10 kHz) led to more packed configurations. Secondly, even for high (negative) $\zeta$ values, a large fraction of the particles tended to be dispersed in the bulk medium. At frequencies higher than 50 kHz, the majority of the particles remained dispersed in the bulk rather than forming planar structures.

\begin{figure}[!h]
  \includegraphics[width=8.3cm]{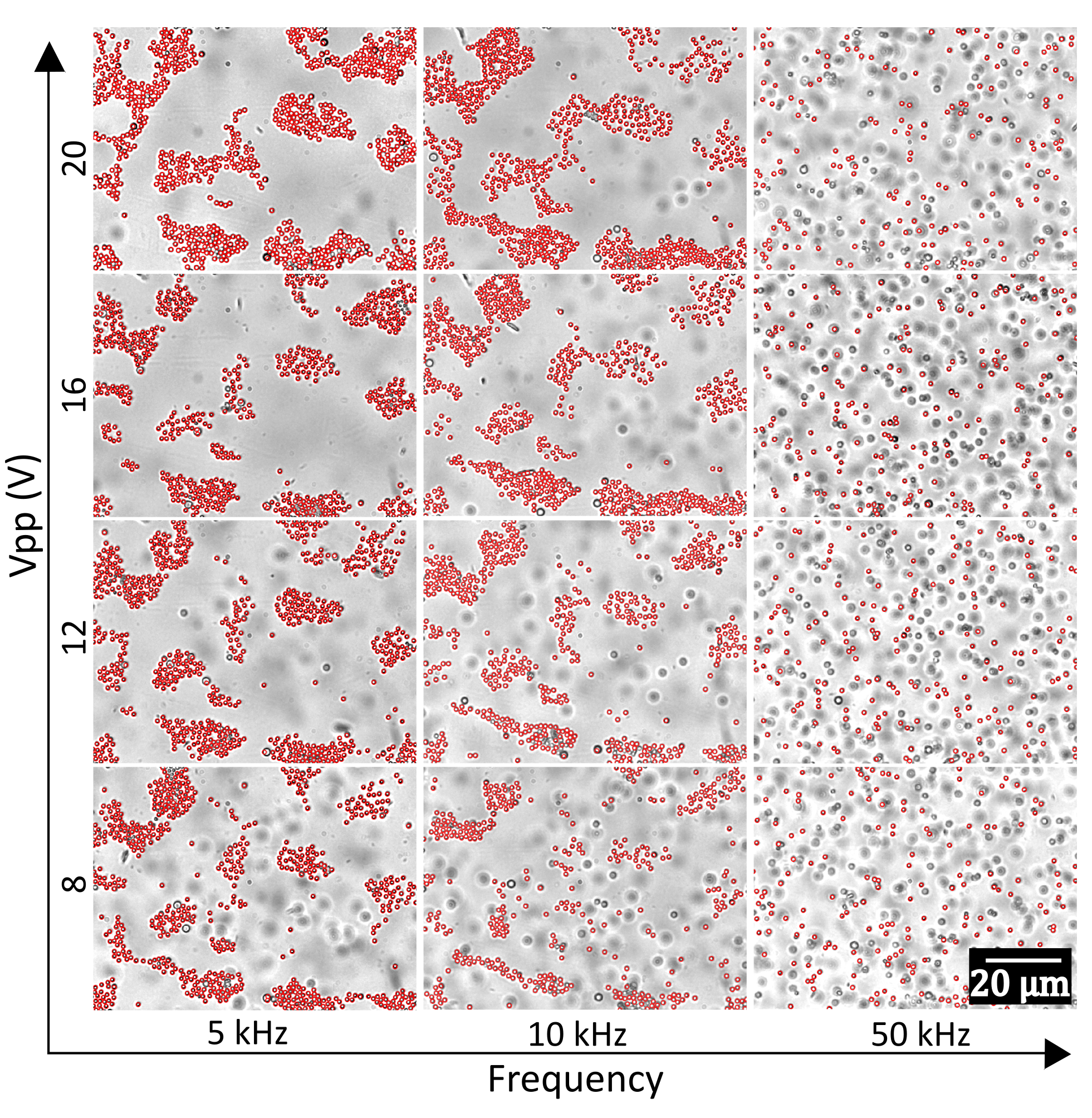}
  \caption{Peak to peak voltage (Vpp) versus frequency optical micrograph of 1 $\mu$m polystyrene spheres functionalised with amine ($\zeta=-94.6$ mV). Cluster size increases with voltage at intermediate frequencies, with no significant effect at low and high frequencies.}
  \label{fig3}
\end{figure}

To quantify the effects of changing $\zeta$ and frequency on the structural organization of the clusters, we constructed Voronoi cell diagrams for each of the images. A representative Voronoi cell diagram of structures formed from 1 $\mu$m polystyrene particles is given in Fig. \ref{fig2}b. For each frequency value, a histogram was plotted for the area of the Voronoi cells followed by a Gaussian fitting (Fig. \ref{fig2}c). The extended tail of the histogram at higher values of the Voronoi cell area was excluded by introducing a window of fitting, in order to ignore the larger cells at the edges and empty areas (Fig. \ref{fig2}c). The value at the peak of the Gaussian was taken as the average value of the cell areas $\left(\langle A \rangle \right)$ for the corresponding frequency, and a plot of $\langle A \rangle $ as a function of frequency was made (Fig. \ref{fig2}d). The nearest-neighbour distances were obtained by calculating the average centre-to-centre distances of the Voronoi cells, and then these distances were averaged $\left( \langle S \rangle \right)$ as before and plotted as a function of frequency (Fig. \ref{fig2}e). $\langle A \rangle $ was observed to increase with the applied frequency as shown in Fig. \ref{fig2}d. In agreement with our observations, at lower frequencies ($\sim$ 5 kHz) particles with higher (negative) values of $\zeta$ showed smaller $\langle A \rangle $ indicating more compact clusters while particles with smaller (negative) $\zeta$ values showed larger $\langle A \rangle $ indicating more spread out clusters (Fig. \ref{fig2}d). The sample with the highest (negative) $\zeta$ (marked in red symbol in Fig. \ref{fig2}d) showed the smallest $\langle A \rangle $. This is in stark contrast to previous observations where the structures were observed to become more close-packed with a decrease in the absolute zeta value.\cite{doi:10.1021/la4048243, Luo_2021} Our observation that more compact structures are formed at higher values of $\zeta$ is unexpected from previous electrohydrodynamic (EHD) calculations, and possible reasons behind this are discussed in section \ref{sectiondiscussions}. There was a general trend of increase in $\langle A \rangle $ with an increasing frequency, indicating that the particles were more spread out in the bulk for all $\zeta$ values. All these observations were also supported by the increase in $\langle S \rangle $ with decreasing $\zeta$ and increasing frequency (Fig. \ref{fig2}e). $\langle A \rangle $ and $\langle S \rangle $ at high frequencies (> 50 kHz) and even at intermediate frequencies (10-50 kHz) deviate from the trend shown in lower frequencies as the particles tend to get increasingly dispersed in the bulk than confined to a plane at these frequencies (Fig. \ref{fig2}d and e).
Other than frequency, the applied voltage also plays a major role in determining the structural configuration as seen in Fig. \ref{fig3} for the case of sample P1A. As the applied voltage is changed from a peak-to-peak value of 8 V to 20 V, the colloidal clusters tend to increase in size. The effect is most pronounced at intermediate frequencies ($\sim$ 10 kHz), whereas there is no significant difference at low and high frequencies. Consequently, $\langle A \rangle $ also shows a gradual decrease with an increase in the peak-to-peak voltage. In summary, size, $\zeta$, frequency and voltage all play a complex and interlinked role in determining the final structure of the colloidal crystals and glasses.

\begin{figure}[!h]
\centering
  \includegraphics[width=8.3cm]{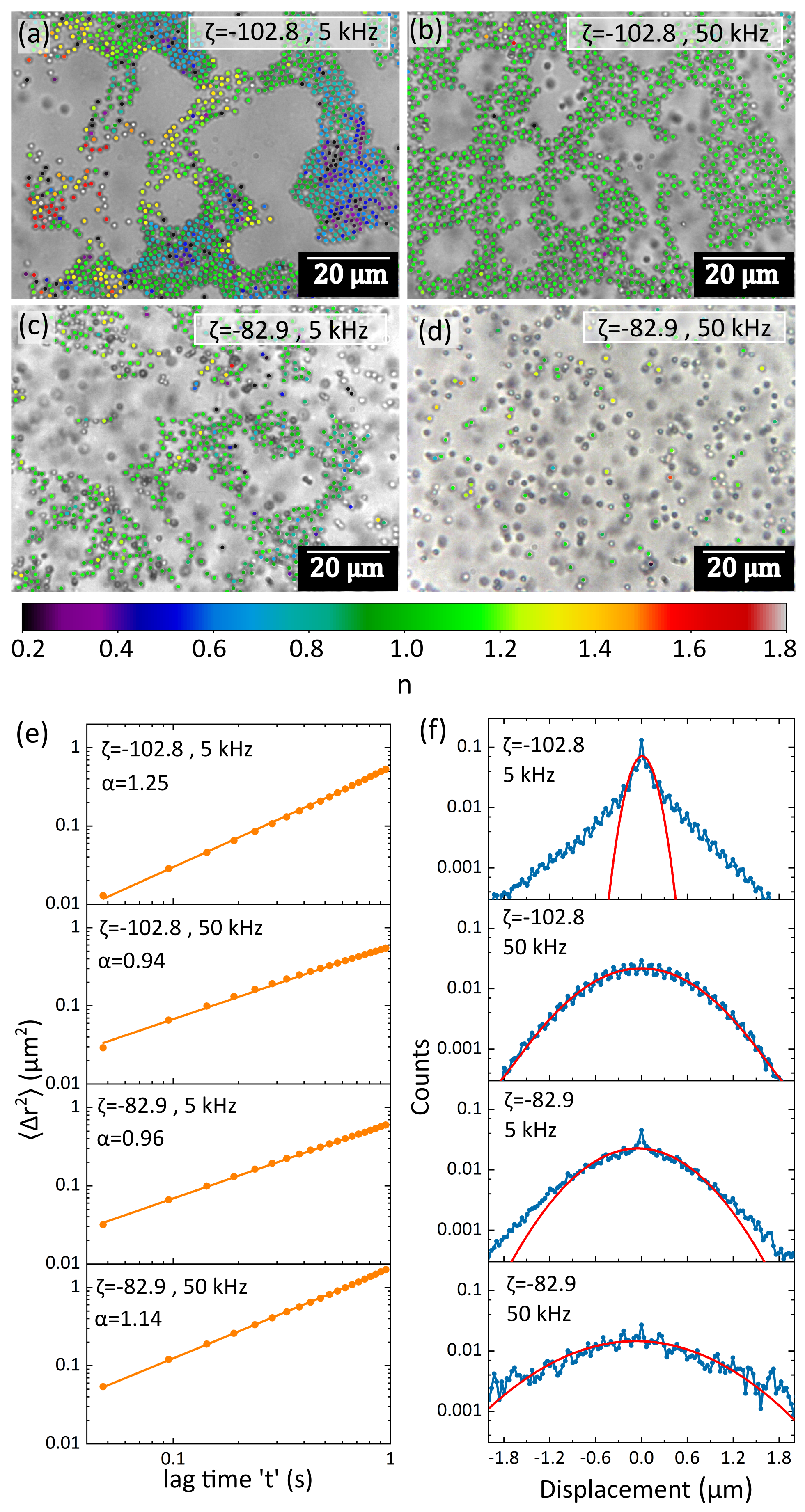}
  \caption{Diffusion dynamics in active colloidal clusters: Colourmapped images showing $n$ values at 20 Vpp for (a) P1X+NaCl at 5 kHz (b) P1X+NaCl at 50 kHz (c) P1X at 5 kHz and (d) P1X at 50 kHz. (e) Log-log plots of time-ensemble averaged \textit{MSD} vs lag time t for P1X+NaCl ($\zeta$ = -102.8 mV) and P1X ( $\zeta$ = -82.9 mV) at 5 kHz and 50 kHz. All the plots show a slope $\alpha$ nearly equal to 1, indicating predominantly Brownian diffusion. (f) Corresponding semi-log displacement distribution plots at 5 kHz and 50 kHz. While at 50 kHz, the displacement distribution is Gaussian, at 5 kHz there is considerable non-Gaussianity for both samples.}
  \label{fig4}
\end{figure}

\subsection{Diffusion dynamics in active clusters}
Palacci \textit{et al.} have shown the formation of light-activated living crystals formed from Janus particles \cite{palacci2013lightlivingcrystals}. In this case, the crystals were called `living' as each constituent colloid was activated by an external optical field and the formation of clusters or crystals was an inherently out-of-equilibrium phenomenon. In our system also, electric field-activated colloids form crystals or glassy structures where they become a constituent part of that dynamic structure. Particles constantly break off and join growing clusters, as the clusters themselves span large areas if allowed to grow for sufficient time. For this reason, we intended to study the dynamical behaviour, that is diffusion and active motion characteristics of the constituent colloids of a large cluster, once the cluster has grown to a considerable size. The particles are expected to show active dynamics, with coupled rotational and translation motion described by the following equations:
\begin{equation}
\dot x=v \cos{\phi}+\sqrt{2D_T}\xi_x,  \dot y=vsin\phi+\sqrt{2D_T}\xi_y, \dot \phi=\sqrt{2D_R}\xi_\phi
\end {equation}
where $x$,$y$ represents the position of the particle, $\phi$ is the orientation, $v$ is the drift velocity of the particle and $\xi_x$,$\xi_y$ and $\xi_\phi$ represent independent white noise stochastic processes, $D_T$ and $D_R$ represent the translational and rotational diffusion constants of the particle. \cite{bechinger} However, once an initial cluster is formed, the inner particles experience a greater caging while the outer particles experience a more open environment. This would lead to different diffusion dynamics among inner and outer particles, thereby directly influencing the morphology of the final structure formed. Chakraborty \textit{et al.} have recently shown that the different local environments experienced by an ensemble of passive colloidal particles in a crowded system, lead to Fickian, yet non-Gaussian diffusion, where the \textit{MSD} remains linear, but the displacement distribution becomes non-Gaussian.\cite{chakrabortyfickianyetnongaussian} We intended to study the dynamics of our field-activated particles which have already become an integral component of a cluster. We expected such a system to show a mixture of pure Brownian motion, subdiffusion, and superdiffusion. Assuming that the characteristic time scale of rotation of our active particles is much larger than the time scale for translational motion, the \textit{MSD} can be calculated using the following equation:\cite{PhysRevLett.99.048102} 
\begin{equation}
   MSD=4D_Tt+v^2t^2 
\end{equation}
where $MSD$ is the mean square displacement, $t$ is the lag time, $D_T$ is the translational diffusion constant and $v$ is the drift velocity of the active particle. Therefore the diffusion exponent \textit{n} for samples with different $\zeta$ values and under different applied frequencies can be calculated using the following equation:
\begin{equation}
    log(MSD)\propto nlog(t)
\end{equation}
where $n$ is the diffusion exponent. For pure Brownian motion, $n$ = 1, while $n$<1 for subdiffusion and $n$> 1 for superdiffusion. From the log-log plot of $MSD$ vs time for each individual particle, we calculated the slope as the diffusion exponent \textit{n} in each case. Fig. \ref{fig4}a-d show the colourmaps for the distribution of $n$ values in two different samples (P1X+ NaCl with $\zeta$= -102.8 and P1X with $\zeta$= -82.9) at 5 kHz and 50 kHz after sufficient time has passed from the application of the field.

The distribution of \textit{n} values was found to be strongly dependent on the applied frequency, indicating different degrees of caging. For example, in the case of P1X + NaCl, there was a wide distribution in the $n$ values at lower frequencies (5 kHz), as seen from Fig. \ref{fig4}a. Particles that were located at the core region of the clusters showed subdiffusion ($n<1$) indicating a high degree of caging from their immediate neighbours. Particles on the outer edge of the clusters, in more sparse regions or singlets showed superdiffusion ($n>1$). In the intermediate regions between these two extremes, we observed pure Brownian motion from the particles as indicated in light green. This is in sharp contrast to the case of 50 KHz (Fig. \ref{fig4}b) where firstly more open clusters were formed, and secondly, \textit{n} was overwhelmingly close to 1 indicating pure Brownian motion with little to no sub or superdiffusion. This shows that at lower frequencies, there is more caging, and more open areas for the particles, leading to the formation of more packed structures. At higher frequencies particles tend to remain relatively more separated from each other, showing predominantly Brownian diffusion, although the applied electric field is present. Similar behaviour was also observed at lower $\zeta$ values: for sample P1X. We again observed a larger distribution of $n$ values at 5 kHz compared to 50 kHz, but the effect was less pronounced than P1X + NaCl. Since even in field-activated systems like ours a large number of particles showed pure Brownian motion, we plotted the \textit{ MSD} vs lag time ($t$) plots for all the particles and calculated the time-ensemble averaged \textit{MSD} $\left( \langle \Delta r^2\rangle \right)$, as well as plotted the displacement distributions. The log-log plots of $ \langle \Delta r^2\rangle$ vs t for all four cases showed a similar overall linear trend (Fig. \ref{fig4}e), with the slope $\alpha$ being very close to 1 indicating predominantly Brownian motion. This indicates that once the crystalline or glassy structures are formed, the dynamics of a large part of the system shifts towards pure Brownian, each constituent particle being locked into the system either physically or due to the electric double layer repulsion from the neighbouring particles. The displacement distributions, however, were more indicative of the system's inner dynamics. For 5 kHz the displacement distributions were distinctly non-Gaussian, the non-Gaussianity being more pronounced for P1X + NaCl (Fig. \ref{fig4}f). For 50 kHz, the displacement distributions were observed to be nearly Gaussian (Fig. \ref{fig4}f). The shift towards pure Brownian dynamics at higher frequencies strongly points to the presence of stronger repulsive barriers among the particles, while lower frequencies lead to areas with stronger and weaker electric field distributions.

\subsection{Frequency induced phase transition in colloidal crystals}
\begin{figure}[!h]
  \includegraphics[width=8.3cm]{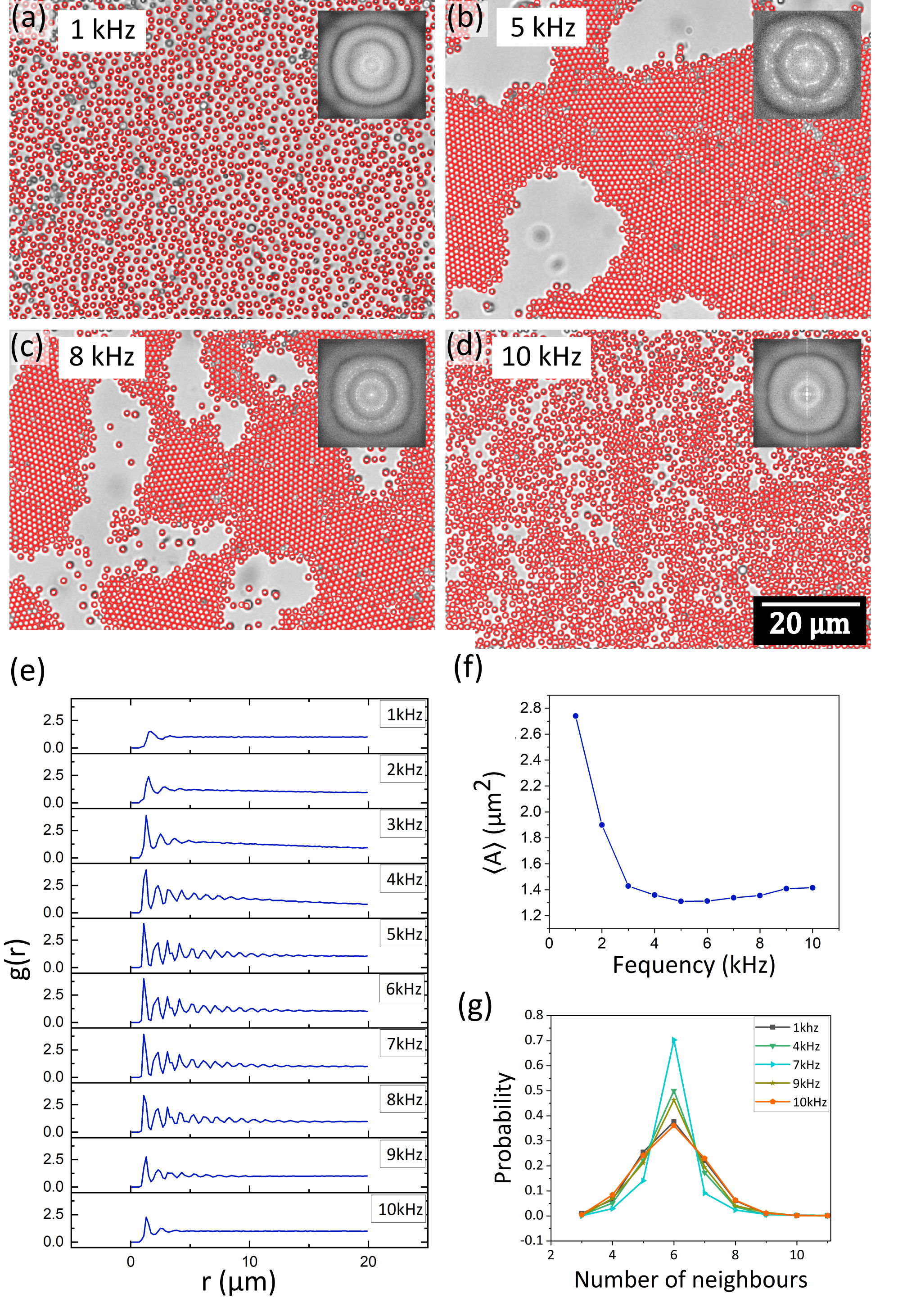}
  \caption{Frequency induced disorder-order-disorder phase transition in 1 $\mu m$ polystyrene particles subject to an applied field frequency of (a) 1 kHz (b) 5 kHz (c) 8 kHz and (d) 10 kHz at 20 Vpp. Inset is the FFT of the image. (e) Radial distribution function $g(r)$ plotted for all frequencies between 1 kHz and 10 kHz shows the creation of order from disorder in the initial stages, followed by a gradual destruction of the order as the frequency is increased. (f) The average Voronoi cell area ($\langle A \rangle$) plotted as a function of frequency showing a drastic decrease followed by a steady increase indicating the most close-packed structure formation at 5 kHz. (g) The probability distribution for the number of nearest-neighbours plotted for frequencies between 1 kHz and 10 kHz indicates a general hexagonal structure but the probability of having six neighbours varies with frequency. }
  \label{fig5}
\end{figure}
The strong dependence of the morphology of the formed structures on the applied field frequency, as well as the different diffusion dynamics at lower and higher frequencies, prompted us to do a detailed structural study across the frequency-induced phase transition from disordered to ordered and back to disordered state of the colloidal crystals. Interestingly, we observed the formation of the most crystalline structure at an intermediate frequency of 5 kHz for 1 $\mu$m polystyrene particles. At lower values of the frequency (1 kHz), destruction of the crystalline order and an increase in the nearest-neighbour distances between the constituent particles were observed (Fig. \ref{fig5}a). From 1 kHz to 5 kHz, the colloids began to crystallize reaching maximum packing and ordering at 5 kHz (Fig. \ref{fig5}b). Beyond 5 kHz, the larger crystals started breaking up into smaller crystallites, as well as individual particles started to break off from the parent crystal (Fig. \ref{fig5}c). Complete destruction of the crystalline order was observed at 10 kHz (Fig. \ref{fig5}d), although the inter-particle distance was observed to be less than the case of 1 kHz, and more bulk suspensions were observed. The fast Fourier transforms (FFT) shown in the inset of the images of Fig. \ref{fig5}a-d show a clear ring pattern superposed with spots for 5 kHz which slowly fades off to a smudged pattern at 10 kHz as well as at 1 kHz, indicating the destruction of the crystalline order. The radial distribution function $g(r)$ plotted in steps of frequency of 1 kHz in Fig. \ref{fig5}e clearly indicates the gradual creation and destruction of the crystalline order as a function of the applied field frequency. At the intermediate frequencies (5-6 kHz), several peaks in the $g(r)$ are observed whereas both at 1 kHz and 10 kHz, a smooth curve is observed with very few peaks. It should be noted here that in this structural reconfiguration, there is a possibility that crystalline order persists, but the packing fraction changes as the frequency is varied. This means that the close-packed crystalline structure becomes more open with a lower value of the packing fraction. To examine this, we again constructed Voronoi diagrams and calculated the Voronoi cell areas. The average Voronoi cell area $(\langle A \rangle)$ showed a minimum near 5 kHz, with a gradual increase between 5 and 10 kHz indicating spreading out of the clusters (Fig. \ref{fig5}f). There is a rapid change in the configuration below 4 kHz where $\langle A \rangle$ rapidly increases and the particles form a planar, glassy structure with large separations from each other. Note that this situation at 1 kHz is drastically different from the case of 10 kHz: in both cases, glassy structures are formed but they are essentially different as indicated by the value of $\langle A \rangle$ in these two frequency regimes (Fig. \ref{fig5}f).  We also observed that at all frequencies, the most probable number of nearest neighbours was 6 (Fig. \ref{fig5}g). However, there is an overall trend of an increase in the probability between 1 kHz and 7 kHz followed by a decrease till 10 kHz. Interestingly, 1 kHz had a slightly higher probability of having six neighbours than 10 kHz. This again implies that at low frequencies, planar, non-close-packed glassy structures are formed, which tend to get packed into large crystalline, close-packed regions at intermediate frequencies, and again pass over to a disordered, bulk suspension at higher frequencies. In sec. \ref{sectiondiscussions} we describe a possible mechanism explaining this frequency-induced phase transition.

\begin{figure}[!h]
\centering
  \includegraphics[width=8.3cm]{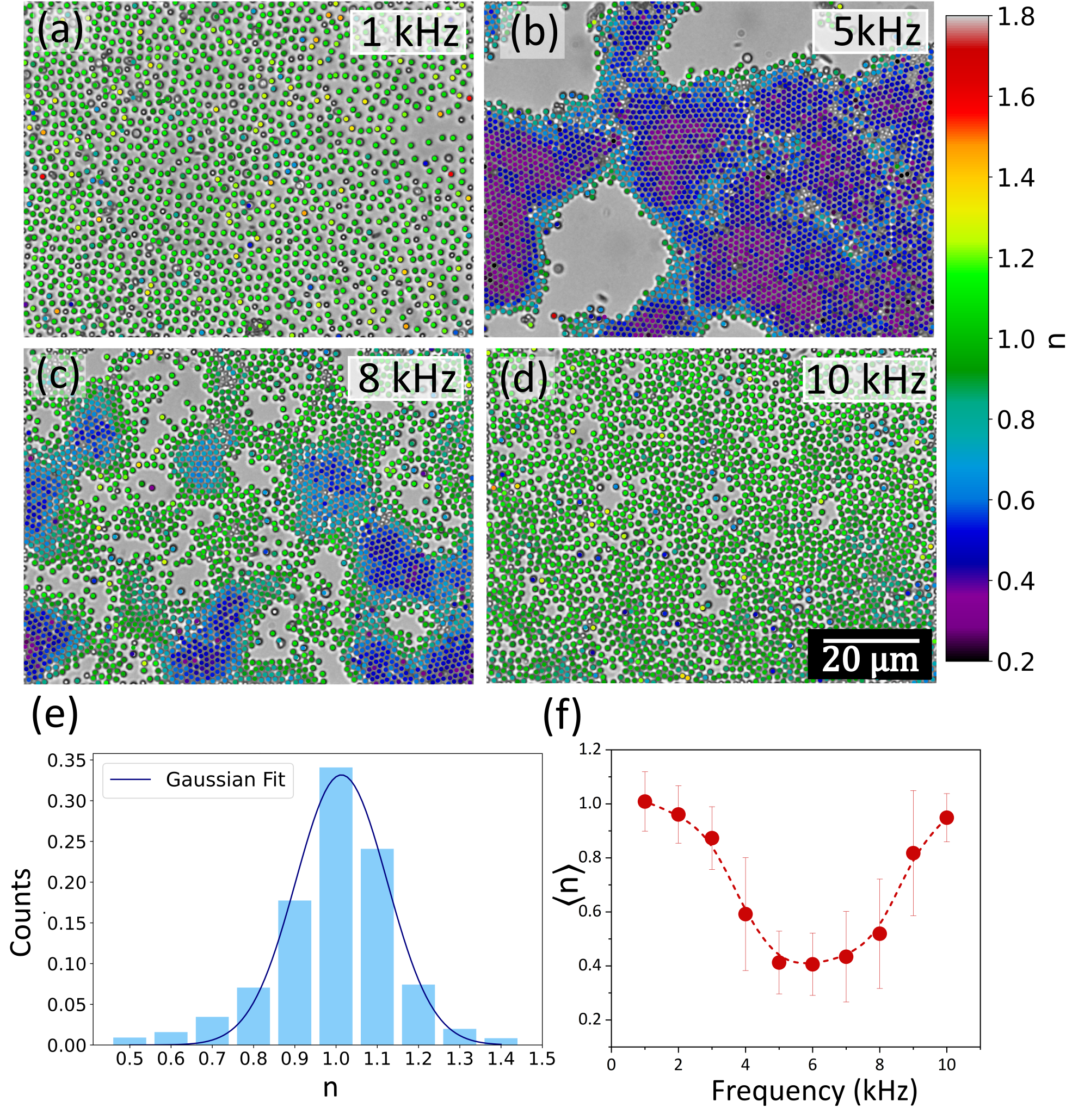}
  \caption{Crystal melting dynamics- Colourmapped images of the diffusion exponent of P1A when subjected to (a) 1 kHz (b) 5 kHz (c) 8 kHz (d) 10 kHz, at 20 Vpp. (e) Representative image of Gaussian fit over the distribution of $n$ values to calculate the mean value $\langle n \rangle$. (f) A plot of $\langle n \rangle$ vs frequency shows that the system is Brownian at low and high frequencies while having both sub and super-diffusive components at intermediate frequencies.}
  \label{fig6}
\end{figure}

It was also observed that multiple frequency sweeps anneal the defects in the growing crystal. Additionally, smaller regions of crystals join with the larger regions, indicating that constant frequency sweeps can give us very large areas of defect-free colloidal crystals. The dynamics of these particles depend strongly on their local environment as shown in Fig. \ref{fig6} a-d. In the interior regions of the large crystalline structures, the diffusion exponent $n$ values are close to 0.5, indicating arrested diffusion due to the presence of neighbouring colloids. On the other hand, the outer regions show $n$ close to 1 or higher, indicating Brownian or superdiffusive behaviours respectively. This indicates that particles in the outer regions have a significant probability to break off from one crystalline zone and join an adjacent growing zone. This was observed multiple times in crystallization experiments. As we go from an ordered to a disordered state at higher frequencies, the diffusion dynamics clearly become more Brownian (indicated by $n$ values close to 1) (Fig. \ref{fig6}d). A calculation of the mean value of the diffusion exponent $\langle n \rangle$ was done by fitting a Gaussian to the $n$ distribution plots for each frequency (Fig. \ref{fig6}e). A plot of the $\langle n \rangle$ values as a function of frequency (Fig. \ref{fig6}f) showed a gradual decrease and a subsequent increase with a minima near 6 kHz. 

A comparison between Fig. \ref{fig5}f (variation of $\langle A \rangle$ with frequency) and Fig. \ref{fig6}f (variation of $\langle n \rangle$ with frequency) indicates the direct interlink between the diffusion dynamics and structural configuration of the colloidal crystals. Both the curves show a minimum near 5-6 kHz indicating arrested diffusion and a greater degree of packing. While $\langle n \rangle$ shows a gradual rise at both high and low-frequency regimes, $\langle A \rangle$ has a steeper response in the low-frequency regime. The increase of $\langle A \rangle$ and $\langle n \rangle$ for the two extreme frequency regimes indicates that the repulsive barrier between the particles tends to dominate locking them into an open, glassy structure and producing predominantly Brownian diffusion dynamics. 

\section{Discussions}
\label{sectiondiscussions}

\begin{figure}[h]
  \includegraphics[width=8.3cm]{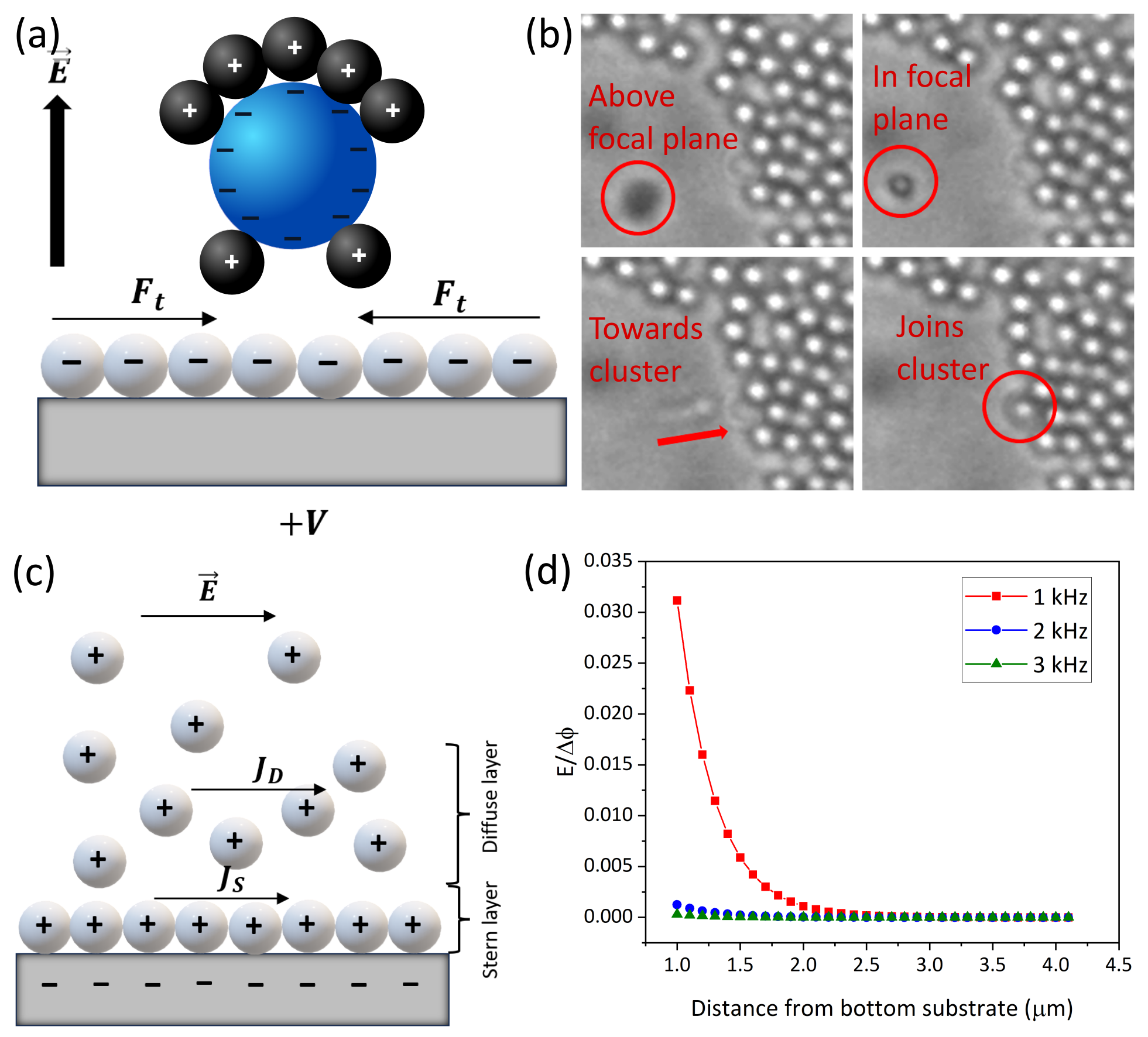}
  \caption{(a) Schematic diagram of a colloidal particle under an applied AC field and situated close to the surface of a polarizable electrode. The tangential component of the local electric field generates a horizontal electrohydrodynamic flow (EHD), bringing particles together to form clusters in a plane perpendicular to the applied field. (b) Bright field images of a particle descending from bulk to the observation plane, and moving horizontally across the surface to join a large cluster. (c) A schematic showing the current $J_S$ due to the Stern layer conductivity and the current $J_D$ due to the diffuse layer conductivity under an applied electric field $\Vec{E}$. (d) Plot showing the rapid increase of the electric field near the surface of an electrode. The effect is more pronounced at lower frequencies (1 kHz) compared to 5 kHz or 10 kHz.   }
  \label{fig7}
\end{figure}

 If we represent the applied AC electric field with frequency $\omega$ as the real part of the complex phasor $\textbf{E}(t) = E_0 e^{i\omega t}\hat{e_z}$, then the time-averaged dipole-dipole interaction force on a particle at position \textbf{r} exerted by another particle residing at the coordinate origin is given as:\cite{katzmeierprl,ningwunonclosepacked}
 \begin{equation}
    \textbf{F}_{dip}(\textbf{r})=\frac{3}{4}\pi\epsilon _m a^2\lvert{C_0}\rvert^2E_{0,rms}^2 \left(\frac{2a}{r}\right)^4 \left[(1-3cos^2\theta)\hat{r}+sin 2\theta\hat{\theta}\right]
\label{fdip}    
\end{equation}

where $C_0$ is the complex dipole coefficient, $\epsilon _m$ is the permittivity of the medium, $a$ is the particle radius, $r$ is the center-to-center distance between the particles, $E_{0,rms}$ is the rms value of the applied electric field and $\theta$ is the angle between the field direction and the line joining the centers of the particles. When $\theta=0^\circ$, the inter-particle force due to dipolar interactions is attractive, and when $\theta=90^\circ$, the force is repulsive. We can clearly see that as the dipolar force varies as $a^6$, the dipolar contribution will dominate for larger particles. This leads to the formation of columns parallel to the field and chains of columns in 3 $\mu$m and 2 $\mu$m particles respectively, in contrast to crystals and glassy structures for 1 $\mu$m particles. The formation of planar structures perpendicular to the field direction for 1 $\mu$m particles implies that the dominant mechanism in this case is not dipolar interactions, but electrohydrodynamic flow (EHD) between the particles and the substrate. This is basically induced charge electro-osmotic (ICEO) flow on the polarizable surface of an electrode. According to Ristenpart \textit{et al.} \cite{ristenpart2004assemblypre}, the tangential velocity $u_t$ due to EHD flow scales as:
\begin{equation}
    u_t \sim \frac{3\varepsilon\varepsilon_0}{\mu\kappa}\frac{(\Delta\phi)^2}{4H^2}\Bigl\{C^\prime_0 + \frac{D\kappa ^2}{\omega} C^{\prime\prime}_0 \Bigr\}
\label{tangentialehdvelocity}    
\end{equation}

where $C^\prime_0$  and $C^{\prime\prime}_0$ are the real and imaginary parts of the dimensionless complex dipole coefficient $C_0$ which is a function of frequency, particle properties like zeta potential and
radius, solution properties like ionic strength, ion valences and mobilities, and the dielectric constant. $\kappa^{-1}$ is the Debye length, $\mu$ is the fluid's shear viscosity, $D$ is the ion diffusivity, $\Delta\phi$ is the applied potential difference between the two parallel electrodes separated by a distance $2H$, $\omega$ is the angular frequency of the applied AC field, $\varepsilon$ is the dielectric constant of the medium and $\varepsilon_0$ is the permittivity of free space respectively.

When $\left[C^\prime_0 + \frac{D\kappa ^2}{\omega} C^{\prime\prime}_0 \right]\ <0$, the EHD flow will be contractile (directed towards the particle), and if $>0$, the flow will be extensile (directed away from the particle). Fig. \ref{fig7}a shows a schematic diagram representing the contractile EHD flow generated due to the force $F_t$ generated on the ions (near the electrode surface) by the tangential component of the local electric field of the particle along with its electric double layer. This force results in horizontal translation of the particles across the surface of the electrode. Even while imaging, the particles were observed to sediment to the bottom electrode and translate horizontally across the electrode surface to attach to a larger cluster (Fig. \ref{fig7}b). At $\zeta = 0$, the effective induced dipole moment of the dielectric particle in a polar solvent like water is directed opposite to the applied field, so the dipole coefficient has a large negative value and a contractile EHD flow is obtained. As $\zeta$ becomes more negative, the concentration of ions around the particles increases, and their surface conductivity comes to play a role, making the dipole moment smaller and less negative. Consequently, the contractile EHD flow becomes smaller. Beyond a certain $\zeta$ value, the particle becomes more polarizable than the medium due to the contribution of the surface conductivity of mobile ions in the diffuse layer, leading to an extensile EHD flow.\cite{yang2019impactsternlayer} An extensile EHD flow will lead to non-close packed structures while a contractile flow will lead to crystal formation. With an increasing (absolute) value of $\zeta$, Woehl \textit{et al.}\cite{doi:10.1021/la4048243} demonstrated a decreasing aggregation rate in colloidal clusters, while Luo \textit{et al.}\cite{Luo_2021} observed a larger separation between between constituent particles in a cluster. In both cases, the zeta potential was varied by using a wide range of electrolytes at different concentrations. In our observations, we however see a completely opposite picture. A higher negative value of $\zeta$ produces more close-packed or crystalline structures while a lower negative value produces more open structures. A possible reason behind this might be the fact that the surface conductivity of the particle consists of not just the diffuse layer conductivity, but also the Stern layer conductivity. The schematic in Fig. \ref{fig7}c shows the two different current densities: $J_S$ due to the Stern layer conductivity and $J_D$ due to the diffuse layer conductivity.\cite{morgan2003ac} The Stern Layer conductivity produces drastically different structures in particles even with nearly similar zeta potentials, for example, Yang \textit{et al.} showed that polystyrene and silica spheres with nearly the same zeta potential ($-$43.8 mV and $-$41.5 mV ) produce non-close packed and crystalline structures respectively due to different Stern layer conductivities \cite{yang2019impactsternlayer}. It was also shown that if the Stern layer conductivity ($\sigma_p$) is high enough, polystyrene particles with $\zeta$ = $-$40 mV ( $\sigma_p$ = 15 pS) will show extensile flow (positive velocity), while particles with $\zeta$ = $-$60 mV ($\sigma_p$ = 100 pS), will show contractile flow, contrary to expectations. Therefore even a higher negative value of $\zeta$ can produce close-packed structures. We suspect that in our case also, the difference in Stern layer conductivities for the different sets of particles has led to an opposite trend in structure formation as a function of $\zeta$. Further theoretical analysis needs to be done to give an exact explanation behind this observed phenomenon.

The disappearance of the crystalline or open structures at high values of the frequency (Fig. \ref{fig2}a) is obvious from the $1/\omega$ frequency dependence in equation \ref{tangentialehdvelocity} (The frequency dependence of $C_0$ is negligible in our relevant frequency range.) As the frequency increases, the tangential velocity of the EHD flow decreases, therefore the particles tend to remain suspended in bulk and are not brought together. On the other hand, at very low frequencies ($\sim$ 1 kHz), the situation resembles the DC scenario more. In this case, the particles neither form columns, nor remain suspended in bulk like the 100 kHz case, but they remain separate on the observation plane. The sedimentation of the particles to the bottom electrode plane is due to the presence of a high electric field at the surface of the polarizable bottom electrode. For the frequency range relevant to us, Ristenpart \textit{et al.} gave the real part of the electric field $E$ near an electrode as:\cite{ristenpart2007electrohydrodynamic} 
\begin{equation}
E=\frac{\Delta\Phi}{2H}\left(1+ \kappa H \left(\frac{\kappa D}{\omega H}\right)^2 \frac{\cosh{(\kappa z)}}{\sinh{(\kappa H)}} \right)e^{-i\omega t} 
\end{equation}
where $\Delta \phi$ is the oscillatory potential difference applied across two parallel planar electrodes separated by a distance $2H$, with the centreline at $z=0$ and the bottom electrode at $z=-H$, $\kappa$ is the inverse of the Debye length, $D$ is the ion diffusivity, and $\omega$ is the applied frequency. A simple plot of $E/\Delta \phi$ as a function of distance from the bottom electrode (Fig. \ref{fig7}d) shows that the electric field is very high near the surface of the electrode at lower frequencies (1 kHz) and has a sharp decrease with distance. For higher frequencies, the electric field even near the electrode surface is much lower. This indicates that the particles will tend to stay close to the surface at lower frequencies and will not come together to form crystalline or open structures (Fig. \ref{fig5}a). This is further emphasized by the observation that the diffusion exponent $n$ value is nearly equal to 1 for a majority of the particles for 1 kHz, indicating, little active motion. In contrast, at a frequency of 5 kHz, a wide distribution of $n$ values is observed, indicating subdiffusion in the core regions and superdiffusion or active motion in the outer regions, and thereby the existence of a substantial amount of EHD flow.

\section{Conclusions}
In summary, we investigated three important aspects of electric field (AC) driven assembly of isotropic colloidal particles: a) the role of several influencing factors including size, zeta potential, frequency and voltage of the applied field in determining the final structural configuration of the colloidal crystals and glasses, b) the dynamics of diffusion and active motion and how that is interlinked with structure formation and c) the frequency driven disorder-order-disorder phase transition in colloidal crystals. 

Initially, we explored the formation of colloidal crystals and glasses under AC electric fields, observing distinct configurations ranging from columnar structures to hexagonal close-packed crystals and planar glassy structures. While larger particle sizes led to the formation of more vertically oriented (parallel to the field) structures, smaller sizes led to in-plane structure formation (perpendicular to the field). Using Voronoi diagram analysis we showed that higher frequencies lead to the formation of more open structures while higher voltages lead to the formation of more close-packed structures. At low frequencies ($\sim$5 kHz), a higher value of the zeta potential leads to close-packed structures, while lower values lead to open structures which is in stark contrast to several previous observations. An examination of the dynamical behaviour of the constituent particles in a colloidal cluster demonstrated a complex interplay of Brownian motion, subdiffusion, and superdiffusion at lower frequencies, which evolved into predominantly Brownian motion at higher frequencies. The variation of the diffusion exponent $n$ indicated different degrees of particle caging and mobility within the clusters at different frequencies. Lastly, we studied the frequency-induced phase transition in colloidal crystals where an open two-dimensional configuration was seen at very low frequencies, and an increase in the frequency led to a gradual development of crystalline order till 6 kHz. Beyond this, the crystalline order began to disappear leading to a completely disordered state at 10 kHz. An analysis of the average diffusion exponent and Voronoi cell area with frequency showed that there is a direct interlink between the nature of diffusion and the structural configuration of the colloidal crystals. Overall, our findings contribute to a systematic, deeper understanding of the complex behaviour of colloidal systems under AC electric fields, with implications for applications in various fields including the fabrication of reconfigurable, smart materials and field-assembled/driven micro-swimmers for targeted drug delivery.

\section*{Author Contributions}
IB and SR fabricated the sample chambers and did the electric field-based experiments. IB performed the data analysis. IB, SR, and IC contributed to the planning of the experiments and writing of the paper. 

\section*{Data Availability}
Most of the data used in this work are either in the main paper or the ESI. Additional data will be available from the authors upon request as most of these are video data and are too bulky to store in a repository.

\section*{Conflicts of interest}
There are no conflicts to declare.

\section*{Acknowledgements}
We acknowledge the Start-up grant (SRG/2021/001696) of the Science and Engineering Research Board (SERB) (at present the Anusandhan National Research Foundation (ANRF)), Government of India and the additional competitive research grant (GOA/ACG/2021-2022/Nov/03) from BITS-Pilani for this work. We also acknowledge the central sophisticated instrumentation facility (CSIF) at BITS-Goa for the zeta-potential measurements.



\balance


\bibliography{rsc} 
\bibliographystyle{rsc} 

\end{document}